\documentclass[]{article}
\usepackage{graphicx}
\usepackage[a4paper,pdftex=true]{geometry}
\usepackage[latin1]{inputenc}
\usepackage[T1]{fontenc}
\usepackage{aeguill}
\usepackage{amssymb}
\usepackage{graphicx}
\usepackage{bbold}
\usepackage{theorem}
\usepackage{fancyvrb}

\theorembodyfont{\upshape}
\newtheorem{theorem}{Theorem}

\newtheorem{proposition}[theorem]{Proposition}

\newtheorem{remark}{Remark}

\newtheorem{definition}{Definition}
\def\proof{\noindent{\bfseries Proof. }}
\def\endproof{\mbox{\ \rule{.1in}{.1in}}}

\newcommand{\VRE}{V^R}
\newcommand{\WRE}{W^R}
\newcommand{\VR}[1]{\VRE\left(#1\right)}
\newcommand{\WR}[2]{\WRE\left(#1,#2\right)}
\newcommand{\VRI}[2]{\VRE\left(#1|#2\right)}

\newcommand{\VE}{V}

\newcommand{\VI}[2]{\VE\left(#1|#2\right)}

\newcommand{\intConf}[2][z]{\oint_{#2}^{#1}}

\newcommand{\partFunc}[3][z]{Z^{#1}_{#2}\left(#3\right)}

\newcommand{\corrFunc}[4][z]{\rho^{#1}_{#2}\left(#3|#4\right)}

\title{$R$-local Delaunay inhibition Model}

\author{Etienne Bertin, Jean-Michel Billiot, Rémy Drouilhet \\
Labsad, Universit{\'e} Pierre Mend{\`e}s France, Grenoble II,\\
1251, avenue centrale,\\
B.P. 47, 38040 Grenoble cedex 9, France.}

\date{}
\begin{document}
\maketitle
 
\centerline{In memory of Etienne Bertin}

\thispagestyle{empty}
\begin{abstract}
Let us consider the local specification system of Gibbs point process with inhibition pairwise interaction acting on some Delaunay subgraph specifically not containing the edges of Delaunay triangles with circumscribed circle of radius greater than some fixed positive real value $R$. Even if we think that there  exists at least a stationary Gibbs state associated to such system, we do not know yet how to prove it mainly due to some uncontrolled ``negative" contribution in the expression of the local energy needed to insert
any number of points in some large enough empty region of the space.
This is solved by introducing some subgraph, called the $R$-local Delaunay graph, which is a slight but tailored modification of the previous one.
This kind of model does not inherit the local stability property but satisfies some new extension called $R$-local stability.
This weakened property combined with the local property provides the existence of Gibbs state.\end{abstract}

\noindent {\bf keywords:} Gibbs states,
Delaunay triangulation, pairwise interaction,
D.L.R. equations, local specifications, correlation functions.

\baselineskip=22pt

\section{Introduction}

There exist many different manners to describe Continuum Gibbs models. One way is using correlation
functions~\cite{Ruelle69,Ruelle70}, another one is rather using local specification~\cite{Preston76}. One could also investigate integral characterization with Palm distribution~\cite{Georgii76,Nguyen76}, empirical measure leading
to ergodic theorem~\cite{Nguyen79}, variational principle, minimizing the excess free energy 
density linking pressure, entropy and energy density~\cite{Georgii94}. In this framework, an important ingredient, particularly useful for the existence of a Gibbs
measure~\cite{Preston76,Ruelle69}, is
the relative compactness assumption taking different form depending on the chosen description (correlation functions, specifications and relative entropy...).
The local stability is a sufficient
assumption for the relative compactness assumption. However this property is very interesting in several other purposes: convergence of Markov chain Monte Carlo (McMC) algorithms to reach equilibrium~\cite{Geyer94,Kendall00}, stochastic FKG domination~\cite{Georgii97},
uniqueness of a Gibbs state via Kirkood-Salsburg equations~\cite{Moraal76,Ruelle69,BBD2}...
One might find interesting to weaken the local stability assumption.
In the classical framework of pairwise interaction Gibbs point process, 
the natural extension is the well-known 
superstable assumption including hard core, inhibition and Lennard Jones pairwise potential.
As already discussed
for example in~\cite{BBD4,BBD3}, this assumption is not well
suited for nearest neighbours models introduced by Baddeley and M{\o }ller~\cite{Baddeley89},
where the neighbourhood relation depends locally on the realization of the process. 
 Kendall {\it et al}~\cite{Kendall99} developed models
generalizing area interaction model called quermass interaction processes and 
intensively studied simulation, statistics, Markov properties (see also \cite{Kendall90,Baddeley96,Moller98}) rather in the 
spirit of non stationary Gibbs states. As it is often the case for nearest-neighbour continuum models,
 such a kind of model is not locally stable .

In \cite{BBD5} we deal with nearest-neighbour continuum Potts model where the repulsion between particles of different type only acts on a Delaunay (sub)graph. This work is an adaptation of the Lebowitz and Lieb soft core continuum Potts model~\cite{Lebowitz72}.
In order to exhibit a phase transition phenomenon for such a model, we mainly need, on the one hand, to use some already known percolation result~\cite{Haggstrom00} on the Delaunay graph and, on the other hand, to choose some Delaunay subgraph for which the existence of the related two-types particles model can be established.
The question arises whether one or several of our existing one-type particle nearest neighbour model could be adapted to this phase transition problem. The spirit of these previous works was to build some Delaunay subgraphs for which the local stability holds. Unfortunately, as a direct consequence, these resulting subgraphs do not behave locally as the original Delaunay graph and do not directly inherit the required percolation property due to some change of connectivity when the number of neighbours dramatically increases.
In order to ensure both the percolation property and the existence of the model, the chosen solution was to define the nearest-neighbour continuum Potts model as a two-types particles Gibbs point process related to some energy function with hard-core component and based on some Delaunay subgraph specifically not containing the edges of Delaunay triangles with circumscribed circle of radius greater than some fixed real value $R$.
Let us point out that, without the hard-core assumption, the proof of the existence of the one-type particle pairwise Gibbs point process based on such a Delaunay subgraph is not yet established even if we additionnaly require the nonnegativeness of the interaction function. However, we are convinced that this model exists. In other words, this simply means that the well-known pairwise inhibition model on the complete graph is not yet adapted to the Delaunay nearest-neighbour framework. The main reason is that
the local energy needed for the insertion of one point is necessarily stable (since obviously nonnegative) for the first one whereas this could not happen for the second one due to some ``negative" residual edges contributing in its expression.
The present study is an attempt to define a first existing version of (pairwise) inhibition nearest-neighbour Gibbs point process.
We investigate to introduce some new Delaunay subgraph, called $R$-local Delaunay. In fact, this is a subgraph of the one considered just above. It mostly preserves the same expected properties and above all its ``local" behavior, that is, the same connectivity at small scale as the Delaunay graph.
Its further characteristic is that no ``negative" contribution of residual edge occurs when inserting any number of points in domain defined as union of balls of radius $R$. As a direct consequence, the related
inhibition nearest-neighbour model inherits some new property, called $R$-local stability,  weaker than local stability but enough to upper bound the correlation functions of the process by some homogeneous Poisson process ones. Combined with the local property intrinsic to such process, this last property provides the existence of stationary Gibbs state.

We may hope that nearest-neighbour continuum models are 
interesting for small temperature (not too small 
for a classic approach) as an alternative of standard 
models on regular networks, because it allows more 
degrees of freedom and may be find  applications in crystallography. 
We may think of the rigidity and plasticity properties of glasses 
or the study of ferromagnetic fluids or 
liquid cristals (smectic A,C, nematic N). 
See for example~\cite{Connelly01,Georgii98} and references therein. 
In particular, it seems that the place of emptyness is important for the study of equilibrium tension in a menbrane~\cite{Connelly01,Menshikov02}.
More generally, it is well-known that 
Voronoi graph and regions (rather called Wigner-Seitz grid and 
Brillouin zone in physics framework) take a fondamental place 
for the understanding of the electrical current, waves propagation 
and phase transitions.

After giving some notations and preliminaries about the $R$-local Delaunay graph in section~\ref{sec:GDL},
we introduce in section~\ref{sec:PPIM} the inhibition pairwise interaction model based on the $R$-local Delaunay graph. After introducing the definition of the $R$-local stability, we establish in section~\ref{sec:ETGM} the existence of a Gibbs measure associated to some local specifications family based on this related energy function.

\section{The $R$-local Delaunay graph}
\label{sec:GDL}
In this section and for the rest of the paper, $R$ designates some fixed nonnegative real number.
For any given Borel set $\Lambda\subset \mathbb{R}^d$, one denotes 
by $\Omega $ and $\Omega_{\Lambda}$ 
the classes of locally finite  subsets of points, called configurations in this paper, in $\mathbb{R}^d$ and $\Lambda$ respectively. In particular, $\Omega_f$  denote the sets of finite configurations in $\Omega $. Moreover, for any set $\Delta$ (not necessarily a Borel set) $\mathcal{P}_2(\Delta)$ designates the set of pairs of points in $\Delta$.
Let $\mathcal{B}$ and $\mathcal{B}_{b}$ be the set of Borel sets and bounded Borel sets of $\mathbb{R}^d $.
 An element $\varphi$ of $\Omega$ could be represented as $\varphi =\displaystyle\sum_{i\in {\bf I\!N}}\delta _{x_i}$ which is a simple counting
Radon measure in ${I\!\!R}^d$ (i.e. all the points $x_i$ of ${I\!\!R}^d$ are
distinct) where $\forall \Lambda \in {\cal B}\,,\,\delta
_{x}(\Lambda )=1_\Lambda(x)$ is the Dirac measure and $1_A(.)$ is the indicator function of a set $A$. 
This space $\Omega$ is equipped with the vague topology, that is to say the weak
topology for Radon measures with respect to the set of continuous functions
vanishing outside a compact set. ${\cal F}$ is the $\sigma $-field
spanned by the maps $\varphi \longrightarrow \varphi (\Delta )\,,\Delta
\in {\cal B}_b$, $\forall\varphi\in \Omega$. The corresponding  $\sigma $-field
${\cal F}_\Lambda$ is similarly defined on $ \Omega_{\Lambda}$ . Furthermore, for any $\Lambda\in{\cal B}_b$,
$$
\left(\Omega,{\cal F}\right)=\left(\Omega_{\Lambda},{\cal F}_{\Lambda}\right)\times
\left(\Omega_{\Lambda^c},{\cal F}_{\Lambda^c}\right)$$  
where $\Lambda^c$ denotes the complementary of $\Lambda$ in ${I\!\!R}^d $. Let  $\widetilde{{\cal F}}_\Lambda$ be the reverse projection of ${\cal F}_\Lambda$
under the previous identification, so that $\widetilde{{\cal F}}_\Lambda$ is a $\sigma$-field on $\Omega$.

A point process $\Phi$ on ${I\!\!R}^d$ (respectively  $\Phi_{\Lambda}$ on $\Lambda$) is a random variable on $\Omega$ (respectively on  $\Omega_{\Lambda}$) and is associated to a probability distribution ${ P}$ on $(\Omega, {\cal F})\,$, (respectively ${ P}_\Lambda$ on $(\Omega_\Lambda,{\cal F}_\Lambda)$).

Some configuration $\varphi$ is said to be in general position when no $d+2$ points lie on the same hypersphere (with no point inside) and no $l+1$ ($l=2,\ldots,d$) points lie on some $l-1$ dimensional affine subspace of $\mathbb{R}^d$.
For any simplex $\psi$ (triangle when $d=2$) in some configuration $\varphi$, one denotes by
$C( \psi ) $ the greatest hypersphere circumscribed by $\psi $ with no point of $\varphi$ inside its interior. The radius and the center (voronoi vertice) of such hypersphere are respectively  denoted by $r(\psi )$ and $c(\psi)$ . One notices that, for any simplex $\psi$, $\#(C(\psi)\cap \varphi)=d+1$ holds only if the configuration $\varphi$ is in general position.

Before defining the $R$-local Delaunay graph  we first need to recall the definition of the Delaunay graph.
\begin{definition}
\label{delaunay} For some  $\varphi$ in $\Omega$ in general position, one defines $Del_{d+1}(\varphi )$ by
the unique decomposition into simplexes $\psi$ in which the convex hull of the hypersphere $C(\psi)$
does not contain any point of $\varphi\setminus \psi  $. \\
The Delaunay graph is then defined by the set of edges~:
$$
Del_2(\varphi )=\bigcup_{\psi\in Del_{d+1}(\varphi )}\mathcal{P}_2(\psi). 
$$    
\end{definition}
According to the previous definition, one can assert in the two dimensional case that the Delaunay graph is a triangulation whenever the configuration $\varphi$ is in general position.  

\begin{figure}[ht]
\begin{tabular}{ll}
\includegraphics[width=5.5cm,height=5.5cm]{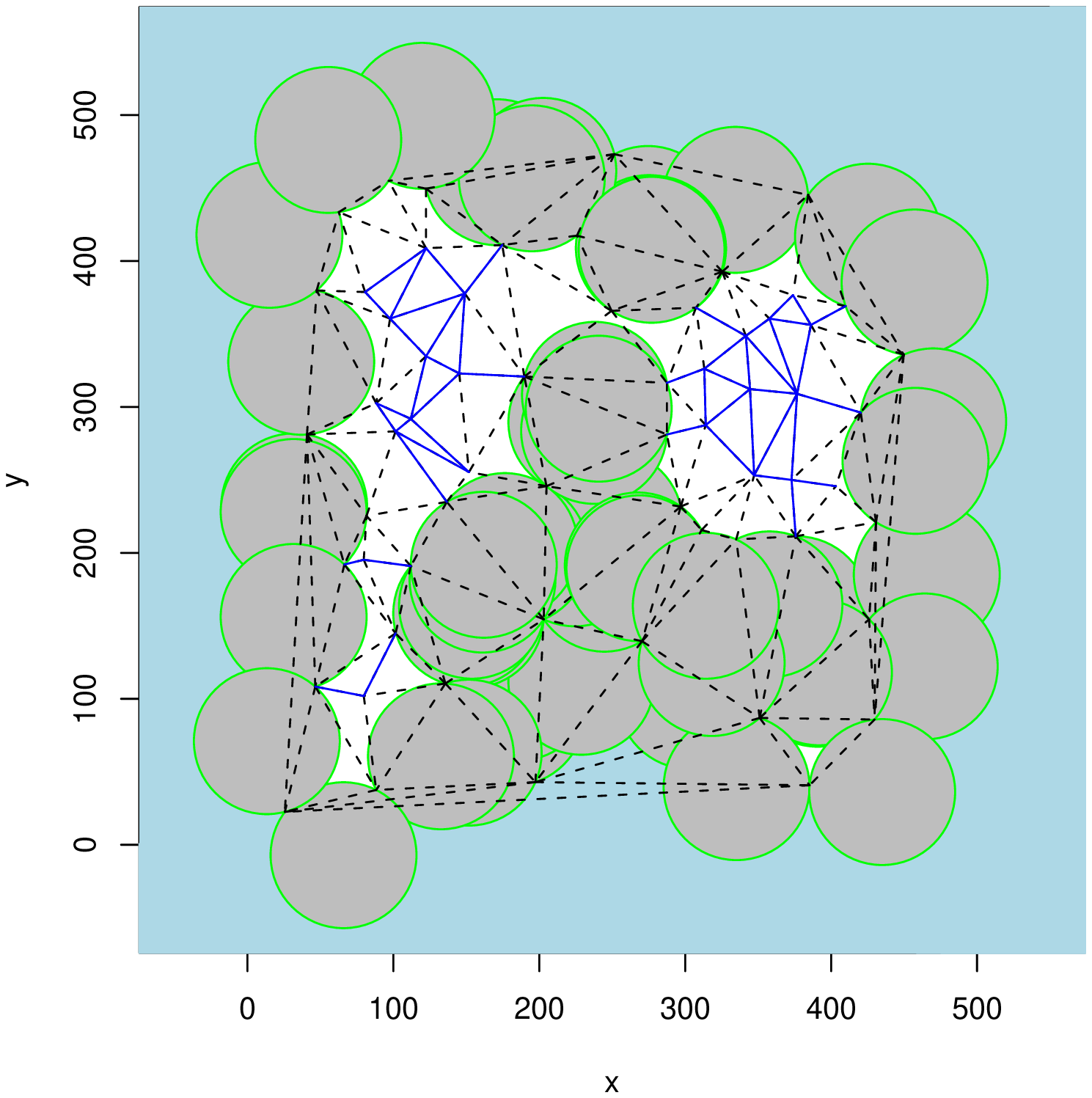}&
\includegraphics[width=5.5cm,height=5.5cm]{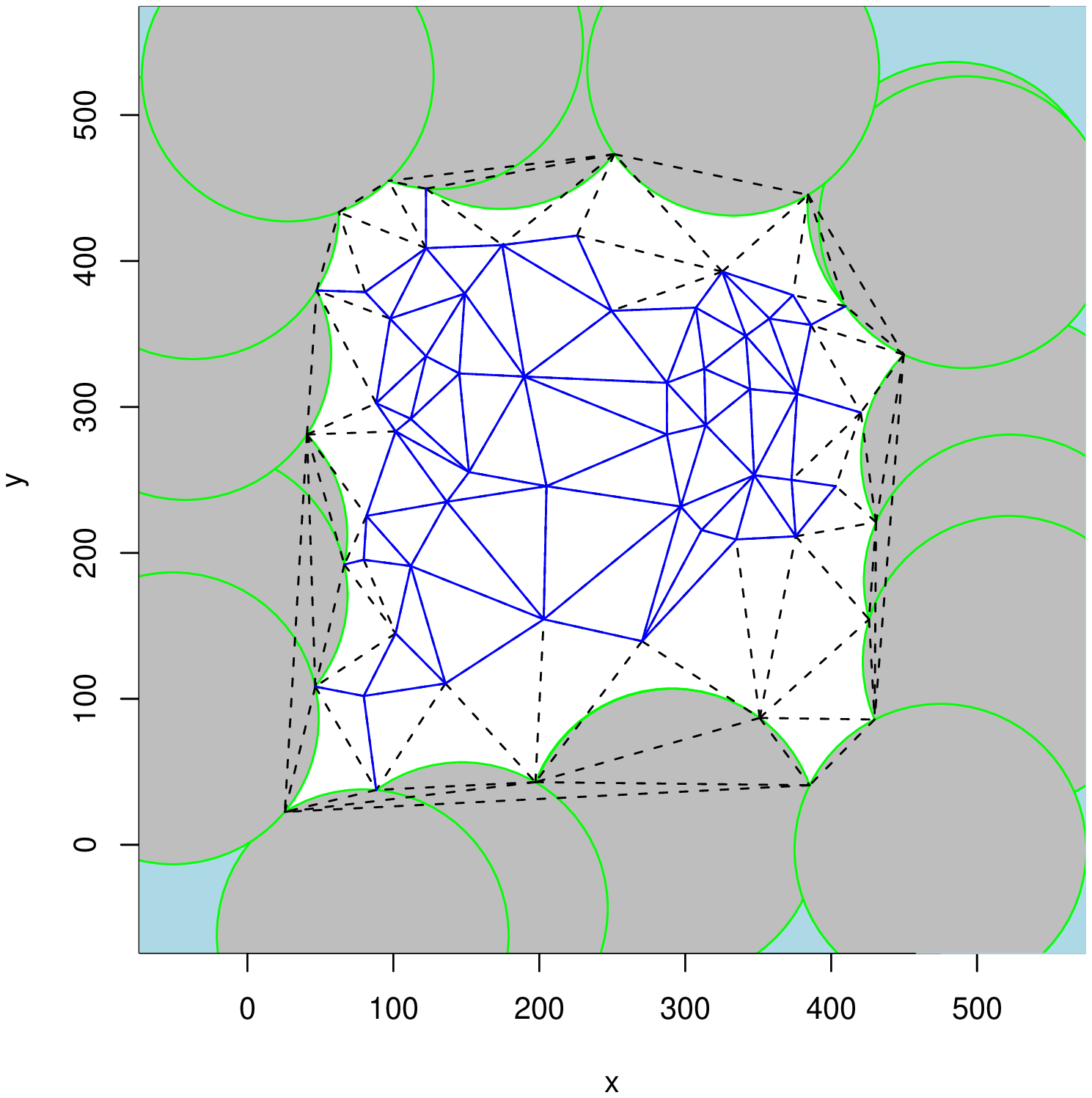}
\end{tabular}
\caption{Both figures represent the same configuration $\varphi$ of some  points (or vertices). However, the first one is given with the R-local Delaunay graph (solid lines) with $R=50$ whereas the second one is for $R=100$. The other edges (dotted lines) are the residual edges of the Delaunay graph. The union of the (darker) gray balls corresponds to the quantity $\nu_R(\varphi)$ defining in some sense the border of the $R$-vacuum. The lighter gray part is the rest of the $R$-vacuum.}
\label{fig-delR50et100}
\end{figure}

Now, we propose to define a subgraph of the Delaunay graph with edges that are particularly of length lower than some positive fixed distance $R$. One need first to introduce, for any set $A$, the set $A\oplus R=\cup_{x\in A}B(x,R)$ where $B(x,R)=\{y\in \mathbb{R}^d:\|x-y\|\leq R\}$ is the usual ball of radius $R$ and $\|x-y\|$ designates the Euclidean distance between the points $x$ and $y$ in $\mathbb{R}^d$. Moreover,  the complement of some set $A$ in $\mathbb{R}^d$ is denoted by $A^c=\mathbb{R}^d\setminus A$.
\begin{definition}
For any  $\varphi\in \Omega$, one defines~:
\begin{enumerate}
\item  \textbf{The $R$-vacuum of $\varphi$}~:
\[
\emptyset_R(\varphi)=\bigcup_{x\in\mathbb{R}^d } \{B(x,R):B(x,R)\cap \varphi= \emptyset\}=\left(\varphi\oplus R\right)^c\oplus R.
\]
\item \textbf{The $R$-local Delaunay graph}~:
\[
Del_2^R(\varphi)=\bigcap_{\psi \in \Omega_f(\emptyset_R(\varphi))}Del_2(\varphi\cup\psi)
\]
where $\Omega_f(\emptyset_R(\varphi))$ is the set of finite configurations in the $R$-vacuum of $\varphi$.
\end{enumerate}
\end{definition}
Thus, one may give further interpretation of this subgraph: any edge of the Delaunay graph of $\varphi$ possibly broken when inserting points in the $R$-vacuum $\emptyset_R(\varphi)$ of $\varphi$ does not lie in the $R$-local Delaunay graph of $\varphi$.  This interpretation leads to another way to define this subgraph. By defining the {\bf influence region} $Z_\varphi(\{x,y\})$ of any edge $\{x,y\} \in Del_2(\varphi)$ by:
\begin{eqnarray*}
Z_\varphi(\{x,y\})&=&\bigcap_{c\in \mathbb{R}^d,r>0}\left\{B(c,r):\{x,y\}\subset \partial B(c,r)\mbox{ and }B(c,r)\cap \varphi\setminus \{x,y\} =\emptyset\right\} \\
&=& \bigcap_{\psi\in Del_{d+1}(\varphi)\,:\,\psi \supset\{x,y\}} B(c(\psi),r(\psi))
\end{eqnarray*}
one derives another characterization of this subgraph:
\[
Del_2^R(\varphi)=\left\{ \left\{x,y\right\}\in Del_2(\varphi):Z_\varphi(\left\{x,y\right\})\cap \emptyset_R(\varphi)=\emptyset \right\}.
\]
Clearly, with respect to this characterization, one may assert some nice property:
\[
Del_2^{R_1}(\varphi)\subset Del_2^{R_2}(\varphi)\mbox{ whenever }R_1\leq R_2
\]
since in this case $\emptyset_{R_1}(\varphi)\supset\emptyset_{R_2}(\varphi)$ (illustrated in figure~\ref{fig-delR50et100}).

Some other properties of the $R$-local Delaunay graph are described in the  next proposition.

\begin{proposition}\label{prop1}
For any configurations $\varphi$, $\varphi_1$ and $\varphi_2$, the following properties holds~:
\begin{enumerate}
\item If $\emptyset_R(\varphi)=\emptyset$ then $Del_2^R(\varphi)=Del_2(\varphi)$.
\item For any $R'>R$, any point $x$ and any borelien set $\Lambda$ , the following holds~:
\[
Del_2^R(\varphi\cap (\Lambda\oplus R)^c) \subset Del_2^R(\varphi).
\]
\item When $\varphi_1$ and $\varphi_2$ are such that $d(\varphi_1,\varphi_2)=\displaystyle\inf_{x_1\in\varphi_1,x_2\in\varphi_2}\|x_1-x_2\|>2R$, 
\[
Del_2^R(\varphi_1\cup \varphi_2)=Del_2^R(\varphi_1)\cup Del_2^R(\varphi_2).
\]
\end{enumerate}

\end{proposition}

\begin{figure}[ht]
\begin{tabular}{ll}
\includegraphics[width=5.5cm,height=5.5cm]{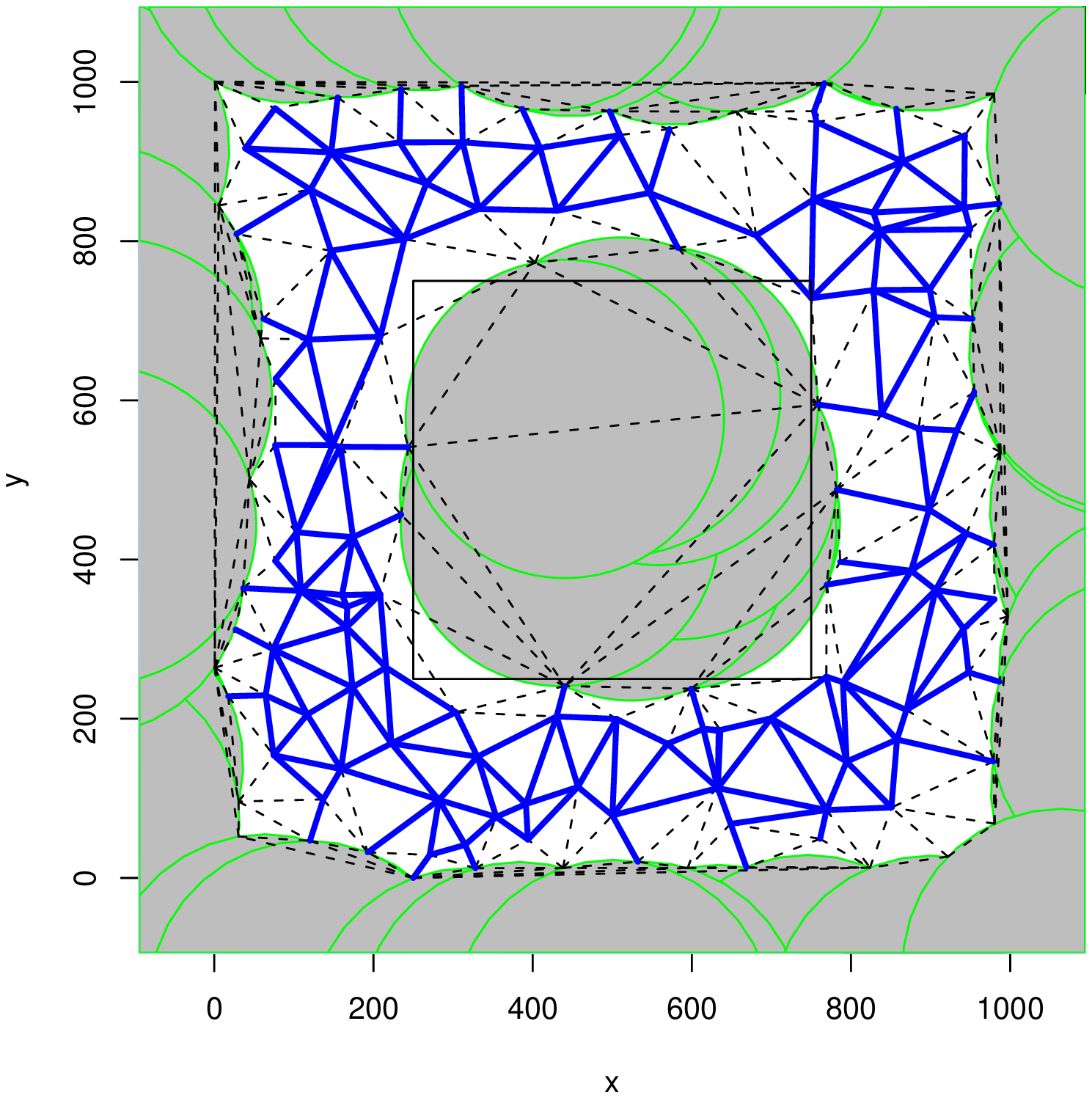}&
\includegraphics[width=5.5cm,height=5.5cm]{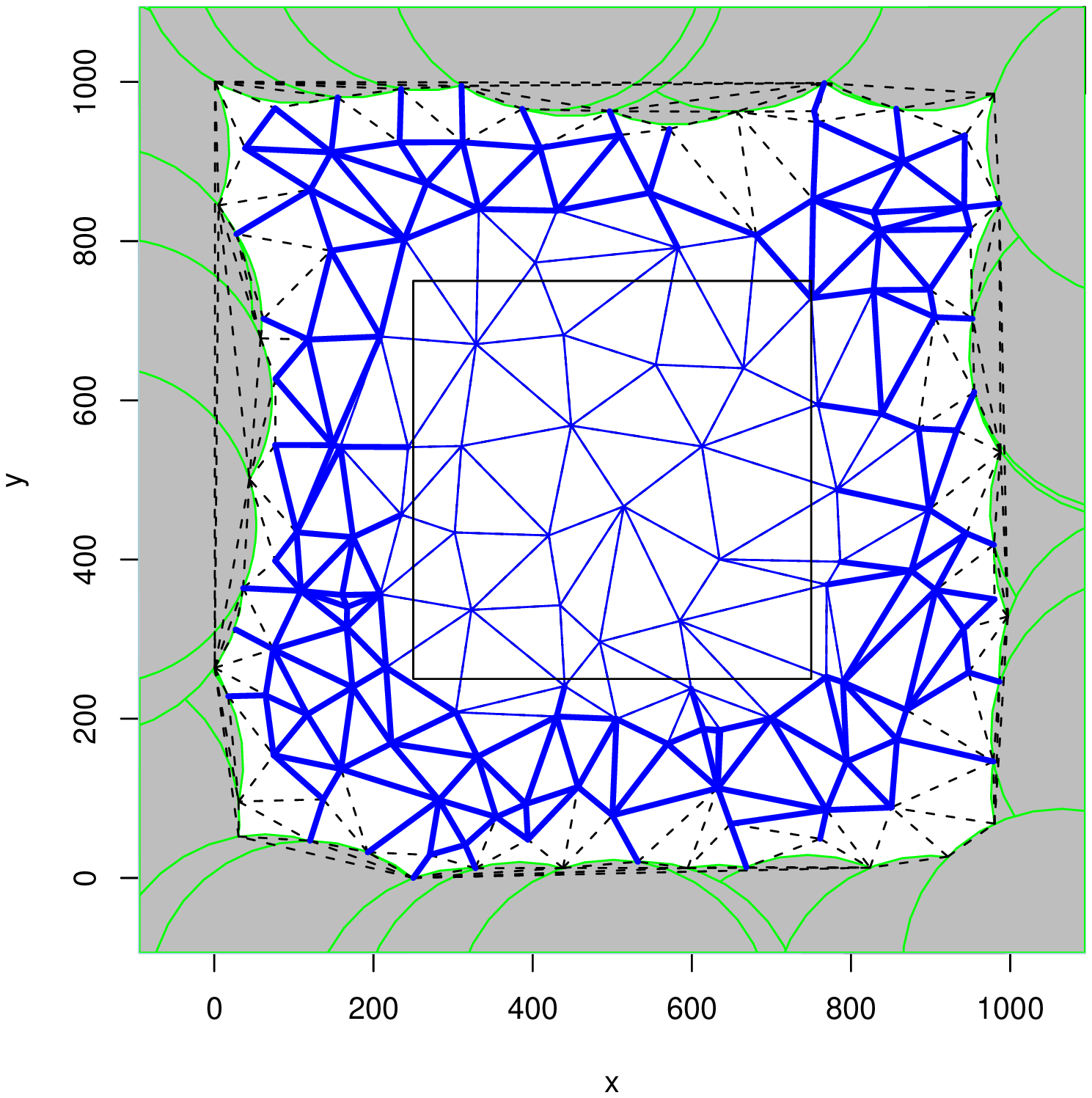}
\end{tabular}
\caption{The figures represent in solid lines the R-local Delaunay graphs ($R=200$) for the configurations $\varphi\cap\Lambda^c$ (on the left) and $\varphi$ (on the right) of some  points where $\Lambda$ is the Borel set bounded by the square. The set of the widest solid edges corresponds to the intersection of both graphs $Del_2^R(\varphi\cap\Lambda^c)$ and $Del_2^R(\varphi)$. In fact, it coincides with the first graph since this graph is included in the second one. One may observe that this is untrue for the Delaunay graphs represented by solid and dotted edges.}
\label{fig-delR1R2}
\end{figure}

The first property means that, for some configuration $\varphi$, the Delaunay graph of $\varphi$ is the same than the $R$-local Delaunay graph of $\varphi$ with $R$ chosen great enough. The second one (see figure~\ref{fig-delR1R2}) is a key-property since it points out that, for this kind of graph, the edges of the graph before the insertion of some point inside a region bigger than  a ball of radius at least equal to $R$  are in the graph  after the insertion. The last one asserts that two subconfigurations of points separated with  a distance greater than $2R$, are disconnected in the graph of the whole configuration. 

Another characteristic property of the $R$-local Delaunay graph is given below.
\begin{proposition}
If any $\{x,y\}\in Del_2(\varphi)$ is such that $d(x,y)=\|x-y\|>2R$, then $\{x,y\}\notin Del_2^R(\varphi)$.
\end{proposition}
\proof
In this case, for any $\psi\in Del_{d+1}(\varphi)$ such that $\{x,y\}\subset\psi$, the radius $r(\psi)$ have to be larger than $R$ \endproof

More generally, for any $ \{x,y\}\in Del_2^R(\varphi)$, we have 
$$diam( Z_\varphi(\{x,y\}))\leq 2R$$
where $diam(\Lambda)= \sup_{(z_1,z_2)\in \Lambda^{2}}\|z_1-z_2\|$ is the diameter of any given Borel set $\Lambda\subset \mathbb{R}^d$.

The first property given in the proposition~\ref{prop1} asserts that the $R$-local Delaunay graph and the Delaunay graph are the same whenever the $R$-vacuum is an empty set. Some kind of generalization of this property is given below by describing the local behavior of the $R$-local Delaunay graph observed on some particular region of $\mathbb{R}^d$ related to the value of $R$.
\begin{proposition}
For any configuration $\varphi$ and any Borel set $\Lambda$ such that~:
\[
\left(\bigcup_{\left\{x,y\right\}\subset\varphi\cap\Lambda}Z_\varphi(\{x,y\})\right)\cap\emptyset_R(\varphi)=\emptyset
\]
which is satisfied whenever $\left(\Lambda\oplus R\right)\cap\emptyset_R(\varphi)=\emptyset$, one may assert that~:
\[
Del_2^R(\varphi)\cap \mathcal{P}_2(\Lambda)=Del_2(\varphi)\cap \mathcal{P}_2(\Lambda).
\] 
\end{proposition}
Clearly, the $R$-local Delaunay graph coincides with the Delaunay graph in region of $\mathbb{R}^d$ with a concentration of points large enough with respect to the value of $R$.

Now, for pratical purposes, we attempt to give an equivalent definition of the $R$-local Delaunay graph by introducing a subregion of the $R$-vacuum. Indeed, in the current form the $R$-local Delaunay graph is not easily computable. We then introduce a special set of edges~:
\[
\mathcal{E}_R(\varphi)=\bigcup_{\psi\in Del_{d+1}(\varphi):r(\psi)\geq R}\, \bigcup_{x\in\psi}\left\{ \psi\setminus\left\{x\right\}:\exists z\in\varphi\setminus\{x\}\mbox{, } r(\left(\psi\setminus\left\{x\right\}\right) \cup\{z\})<R \right\}.
\]
For any $\xi\in\mathcal{E}_R(\varphi)$, let us denote by $c_R(\xi)$ the unique point such that~:
\[
\left(\forall x\in \xi,\|x-c_R(\xi)\|=R\right)\mbox{ and }B(c_R(\xi),R)\cap(\varphi\setminus\xi)=\emptyset.
\]
We propose to define some sort of border of the $R$-vacuum $\emptyset_R(\varphi)$ of some configuration $\varphi$ by introducing the subset  $\mathcal{V}_R(\varphi)$ of $\emptyset_R(\varphi)$~:
\[
\mathcal{V}_R(\varphi)=\bigcup_{\xi\in\mathcal{E}_R(\varphi)}B(c_R(\xi),R).
\]
Consequently, the $R$-vacuum can be decomposed into two parts (not necessarily disjoint)~:
\[
 \emptyset_R(\varphi)=\mathcal{V}_R(\varphi)\cup\left(\bigcup_{\psi\in Del_{d+1}(\varphi):r(\psi)\geq R}B(c(\psi),r(\psi))\right)
\] 
This is illustrated by the previous figures in the two-dimensionnal case.

By applying the following proposition, it is almost easy to propose an algorithm in order to compute the $R$-local Delaunay graph for some configuration $\varphi$. 
\begin{proposition}
The residuals edges in the difference between the Delaunay graph and the $R$-local Delaunay graph  can be decomposed into the union of two disjoint sets~:
\[
Del_2(\varphi)\setminus Del_2^R(\varphi)= \mathcal{R}es_1^R(\varphi)\cup\mathcal{R}es_2^R(\varphi)
\]
where 
\[
\mathcal{R}es_1^R(\varphi)=\bigcup_{\psi\in Del_{d+1}(\varphi)}\left\{\xi\in\mathcal{P}_2(\psi):r(\psi)\geq R\right\}
\]
and 
\[
\mathcal{R}es_2^R(\varphi)=\left\{\xi\in Del_2(\varphi)\setminus\mathcal{R}es^R_1(\varphi):Z_\varphi\left(\xi\right)\cap\mathcal{V}_R\left(\varphi\right)\neq\emptyset \right\} 
\]
\end{proposition}

\begin{remark}
The same idea can be applied to some classical subgraphs of the Delaunay graph like the Gabriel and the Relative Neighbours graphs. Recall that these subgraphs are respectively defined as follows:
\[ 
G(\varphi)=\left\{ \left\{x,y\right\}\in Del_2(\varphi): B(\frac{x+y}{2},\frac{ \|x-y\|}{2}) \cap \varphi\setminus \{x,y\} =\emptyset\right\} 
\]
and 
\[ 
RNG(\varphi)=\left\{ \left\{x,y\right\}\in Del_2(\varphi):  B(x,\|x-y\|)\cap B(y,\|x-y\|) \cap \varphi\setminus \{x,y\} =\emptyset\right\}. 
\]
Indeed, we have just to define the influence regions  $Z_\varphi^{G}(\{x,y\})$ and $Z_\varphi^{RNG}(\{x,y\})$ of any edge $\{x,y\}$ of each graph respectively  by~:
\[
Z_\varphi^{G}(\{x,y\})=B(\frac{x+y}{2},\frac{ \|x-y\|}{2})
\] 
and 
\[
Z_\varphi^{RNG}(\{x,y\})= B(x,\|x-y\|)\cap B(y,\|x-y\|).  
\]
One may derive one another characterization of these new $R$-local subgraphs by asserting~:
\[
G^{R}(\varphi)=\left\{ \left\{x,y\right\}\in G(\varphi):Z_\varphi^{G}(\left\{x,y\right\})\cap \emptyset_R(\varphi)=\emptyset \right\}.
\]
and
\[
RNG^{R}(\varphi)=\left\{ \left\{x,y\right\}\in RNG(\varphi):Z_\varphi^{RNG}(\left\{x,y\right\})\cap \emptyset_R(\varphi)=\emptyset \right\}.
\]
\end{remark}

\section{Inhibition interaction model on the $R$-local Delaunay graph}
\label{sec:PPIM}

We first introduce the definition of the energy function induced by the $R$-local delaunay graph.
\begin{definition}
Given any fixed $R>0$, one defines the $R$-energy of some finite configuration $\varphi$ by~:
\begin{equation}\label{def-VR}
\VR{\varphi}=\sum_{\{x,y\}\in Del_2^R(\varphi)}\phi(\{x,y\}) 
\end{equation}
where $\phi$ is some upper bounded nonnegative pairwise interaction function.
\end{definition}
The previous model could be easily extended by adding interaction terms function of all order. In the rest of this paper, we only deal with the pairwise interaction but all the results remains valid for these extensions whenever the interaction functions of all order are nonnegative and upper bounded.

\begin{remark}
Another similar model is the one with pairwise interaction between Voronoi vertices. In fact, each influence region of Delaunay edge is the intersection between two Delaunay disks for edge inside the convex hull of points and just one Delaunay disk for edges belonging to the convex hull. The local Delaunay graph then take into account interaction between Voronoi vertices and some points characterizing the R-vacuum region. By denoting ${\cal V}(\varphi)=\left\{  c(\psi)\right\}_{\psi\in Del_3(\varphi)}$ the set of the Voronoi vertices
we define 
\[
{\cal V}_{2}^{R}(\varphi)={\Bigl \{ } \{c(\psi_{1}),c(\psi_{2})\}\in { \Bigl(}{\cal V}(\varphi){\Bigr)}^{2}: \#(\psi_{1}\cap \psi_{2})=2,\quad r(\psi_{1})\leq R, \quad r(\psi_{2})\leq R ,$$
$$ B(c(\psi_{1}),r(\psi_{1}))\cap B(c(\psi_{2}),r(\psi_{2}))\cap B(c_{R}(\xi),R)= \emptyset, \forall \xi\in \mathcal{E}_{R}(\varphi){\Bigr \} }.
\]
The following finite energy is of the same kind of the $R$-energy:
\[
V_R(\varphi)=\sum_{\{c(\psi_{1}),c(\psi_{2})\}\in{\cal V}_{2}^R(\varphi)}\phi(\{c(\psi_{1}),c(\psi_{2})\})
\]
where $\phi$ is some upper bounded nonnegative interaction function between Voronoi vertices.
\end{remark}

As usually, the mutual energy and  the conditional energy between two configurations $\varphi$ and $\psi$ are respectively  defined by~:
\[
\WR{\varphi}{\psi}=\VR{\varphi\cup\psi}-\VR{\varphi}-\VR{\psi}
\]
and 
\[
\VRI{\varphi}{\psi}=\VR{\varphi\cup\psi}-\VR{\psi}=\VR{\varphi}+\WR{\varphi}{\psi}.
\]

Due to the properties of the $R$-local Delaunay graph, we have the  following result.

\begin{proposition}\label{prop-VR}
\begin{enumerate}
\item For any set $\Delta$, one has~:
\begin{equation}
\VRI{\varphi_{\Delta\oplus R}}{ \varphi_{(\Delta\oplus R)^c}}=\sum_{\{x,y\}\in Del_2^R(\varphi)\setminus Del_2^R(\varphi_{(\Delta\oplus R)^c}) }\phi(\{x,y\}) \label{eq-VRI}
\end{equation}
and
\[
\WR{\varphi_{\Delta\oplus R}}{\varphi_{(\Delta\oplus R)^c}}=\sum_{\{x,y\}\in Del_2^R(\varphi)\setminus (Del_2^R(\varphi_{(\Delta\oplus R)^c})\cup Del_2^R(\varphi_{\Delta\oplus R}) }\phi(\{x,y\}).
\]
\item If $\Delta_1$ and $\Delta_2$ are two Borel sets such that $d(\Delta_1,\Delta_2)>2R$, the following holds~:
\[
\WR{\varphi_{\Delta_1}}{\varphi_{\Delta_2}}=0.
\]
\item \textbf{Finite range property:} for any bounded Borel set $\Delta$, 
\begin{equation}
\VRI{\varphi_\Delta}{\varphi_{\Delta^c}}=\VRI{\varphi_\Delta}{\varphi_{\Delta\oplus6 R \setminus\Delta}}.\label{eq-FR}
\end{equation}
\end{enumerate}
\end{proposition}
\proof
In fact because of the translation invariance property of the local energy,
it is sufficient to prove that for any  $\varphi\in \Omega$,
$$
\VRI{0}{\varphi}=\VRI{0}{\varphi_{6R}}.
$$
where, for shortness, one denotes $ \varphi_{6R}=\varphi_{B(0,6R)}$.
Roughly speaking, let us show that

$$ Del_2^R(\varphi){\bf \Delta} Del_2^R(\varphi\cup\{ 0\} )=Del_2^R(\varphi_{6R}){\bf \Delta}
  Del_2^R(\varphi_{6R}\cup\{ 0\} )$$

where ${\bf \Delta}$ denotes the symmetric difference operator
 (i.e. for two sets $A$ and $B$, $A {\bf \Delta} B=(A\setminus B) \cup( B\setminus A)$).

Clearly as one remarks after proposition 4, the diameter of the influence region for any edge of the $R$-local Delaunay graph is bounded by $2R$.

 So one has for any $\xi=\{x,y\}\in Del_2^R(\varphi){\bf \Delta} Del_2^R(\varphi\cup\{ 0\} )$ (resp. $\xi=\{x,y\}\in Del_2^R(\varphi_{6R}){\bf \Delta} Del_2^R((\varphi_{6R}\cup\{ 0\} ))$):
$$ Z_\upsilon(\xi)\subset B(0,4R)$$
where $\upsilon$ can be chosen as $\varphi$ or $ \varphi\cup\{ 0\} $ (resp. $\varphi_{6R} $ or $\varphi_{6R}\cup\{0\}$).
Moreover if one notices that 
$$\emptyset_R(\varphi)\cap B(0,4R)=\emptyset_R(\varphi_{6R})\cap B(0,4R)$$
and  $$\emptyset_R(\varphi\cup\{0\} )\cap B(0,4R)=\emptyset_R(\varphi_{6R}\cup\{0\}))\cap B(0,4R)$$
it follows by the definition of the $R$-local Delaunay graph that 
\begin{eqnarray*}
Del_2^R(\varphi){\bf \Delta} Del_2^R(\varphi\cup\{0\} )&=& Del_2^R(\varphi){\bf \Delta} Del_2^R(\varphi\cup\{0\})\cap \mathcal{P}_2(B(0,4R))\\
&=&Del_2^R(\varphi_{6R}) {\bf \Delta}Del_2^R(\varphi_{6R}\cup\{0\}) \cap \mathcal{P}_2(B(0,4R))\\
&=& Del_2^R(\varphi_{6R}) {\bf \Delta}Del_2^R(\varphi_{6R}\cup\{0\})       
\end{eqnarray*}
and the proof is complete \endproof

Defined as a subgraph of the Delaunay graph, the $R$-local Delaunay graph is linear in the planar case and then the $R$-energy $\VRE$ inherits of the stability property. In the higher dimensional case, stability occurs since the interaction function is assumed to be nonnegative.

Furthermore, one can notice that the $R$-energy is not locally stable due to its local behavior as the Delaunay graph on each region of space with high enough density of points. This present work is then really different from the previous ones \cite{BBD3,BBD4} where the goal was to build some subgraphs of the Delaunay graph providing local stability.

\section{The $R$-local stability}

However, one may assert some new property for the $R$-energy $\VRE$ which is an extension of the local stability.
\begin{definition}
For some nonnegative real number $R$,  an energy function $\VE$ is said to be $R$-locally stable if there exists $K\geq 0$ such that for any finite configurations $\varphi$, $\varphi_1$ and $\varphi_2$, and any subset $\Delta$ satisfying $\Delta \supset \varphi$,$\varphi_1\subset\Delta\oplus R$ and $\varphi_2\subset(\Delta\oplus R)^c$:
\begin{equation}
\VI{\varphi\cup \varphi_1}{\varphi_2} \geq -K  \#(\varphi\cup\varphi_1) \label{eq-RLS}
\end{equation}
\end{definition}
One may arrange the $R$-local stability as some property between the stability (acting globally) and the local stability.
Indeed, on the one hand when $R$ vanishes and $\Delta=\varphi$, the $R$-local stability is similar to the local stability and on the other hand when $\Delta=\mathbb{R}^d$ (i.e. $\varphi_2=\emptyset$) the $R$-local stability coincides with the (global) stability.

\begin{proposition}\label{prop-RLS}
The $R$-energy $\VRE$ with nonnegative interaction function is $R$-locally stable.
\end{proposition}
\proof This is a direct consequence of the property~(\ref{eq-VRI}) with $K=0$
\endproof

This property will play later some key-role in the proof of the non emptyness of stationary Gibbs state based on the $R$-energy.

Given any outside  configuration $\varphi^o$, Ruelle \cite{Ruelle69} has introduced the following quantity:
\begin{equation}\label{eq-corrFunc}
\corrFunc{\Delta,\Lambda}{\varphi}{\varphi^o}:=\frac{z^{\#\varphi}}{\partFunc{\Lambda}{\varphi^o}}\intConf{\Lambda\setminus \Delta} d\psi e^{-\VRI{\varphi \cup\psi}{\varphi^o_{\Lambda^c}}}
\end{equation}
where for any measurable function $f$~:
 \[ 
\intConf{\Lambda} d\varphi f(\varphi)=\sum_{n=0}^{+\infty}\frac{z^n}{n!}\int_{\Lambda^n}dx_1\ldots dx_n f(\underbrace{x_1,\ldots,x_n}_{\varphi}).
\]
As a particular case, $\Delta=\emptyset$, one can derive the correlation function $\corrFunc{\Lambda}{\varphi}{\varphi^o}:=\corrFunc{\emptyset,\Lambda}{\varphi}{\varphi^o}$, satisfying
\[
\corrFunc{\Delta,\Lambda}{\varphi}{\varphi^o} \leq \corrFunc{\Lambda}{\varphi}{\varphi^o}.
\]
An interesting well-known property is then to  prove that this correlation function are upper bounded by the correlation function of some Poisson process, that is, of the form~$\xi^{\#\varphi}$.

\begin{proposition}\label{prop-corrFunc}
If some energy function $V$ is $R$-locally stable then
\[
\corrFunc{\Lambda}{\varphi}{\varphi^o}\leq \xi^{\#\varphi}\mbox{ where }\xi=z e^K e^{ze^K|B(0,R)|}.
\]
\end{proposition} 
\proof For any $\Delta$ and $\Lambda \supset \Delta\oplus R$,
\begin{eqnarray*}
\corrFunc{\Lambda}{\varphi}{\varphi^o}&=&\frac{z^{\#\varphi}}{\partFunc{\Lambda}{\varphi^o}}\intConf{\Lambda} d\psi \, e^{-\VI{\varphi\cup\psi}{\varphi^o}}\\
&=&\frac{z^{\#\varphi}}{\partFunc{\Lambda}{\varphi^o}}\intConf{\Lambda\cap(\Delta\oplus R)} d\psi_1\intConf{\Lambda\cap (\Delta\oplus R)^c} d\psi_2 \, e^{-\VI{\varphi\cup\psi_1}{\psi_2\cup \varphi^o}-\VI{\psi_2}{\varphi^o}}\\
&\leq&\frac{(ze^K)^{\#\varphi}}{\partFunc{\Lambda}{\varphi^o}}\intConf{\Lambda\cap (\Delta\oplus R)}d\psi_1 e^{K\#\psi_1}\intConf{\Lambda\cap(\Delta\oplus R)^c} d\psi_2 \, e^{-\VI{\psi_2}{\varphi^o}} \mbox{\hspace*{.5cm}(by $R$-local stability)}\\
&\leq&\left(ze^K\right)^{\#\varphi}e^{ze^K|\Lambda\cap(\Delta\oplus R)|}\frac{\partFunc{\Lambda\cap(\Delta\oplus R)^c}{\varphi^o}}{\partFunc{\Lambda}{\varphi^o}}\\
&\leq& \left(ze^K\right)^{\#\varphi}e^{ze^K|\Lambda\cap(\Delta\oplus R)|} \mbox{\hspace{1cm} (since }\partFunc{\Lambda\cap(\Delta\oplus R)^c}{\varphi^o}\leq\partFunc{\Lambda}{\varphi^o}\mbox{)}
\end{eqnarray*}
Finally, applying this result when $\Delta=\varphi$, this leads to
\[
\corrFunc{\Lambda}{\varphi}{\varphi^o}\leq \left(ze^Ke^{ze^K|B(0,R)|}\right)^{\#\varphi}
\]
since $|\Lambda\cap(\Delta\oplus R)|$ is then upper bounded by $\#\varphi \times |B(0,R)|$\endproof

\section{Existence of a Gibbs measure based on the $R$-energy}
\label{sec:ETGM}

At this stage, everything that one needs in order to prove the existence of a Gibbs measure related to the energy function $\VRE$, was already introduced. 
We then first recall the definition of local specifications based on the $R$-energy.

\begin{definition}
Given any outside  configuration $\varphi^o$, the following family of measures\\ $\Pi=\{\Pi_\Lambda(\cdot,\omega)\}_{\Lambda\in \mathcal{B}_{b}}$ on $(\Omega,\mathcal{F})$ is a system of local specifications~: 
\[
\forall F\in\mathcal{F},\quad \Pi_\Lambda(F|\varphi^o)=\frac 1{Z_\Lambda(\omega)} \intConf{\Lambda} d\varphi \, e^{-\VRI{\varphi}{\varphi^o_{\Lambda^c}}}\mathbb{1}_F(\varphi\cup\varphi^o_{\Lambda^c})
\]
where the partition function is given by $\displaystyle\partFunc{\Lambda}{\varphi^o}= \intConf{\Lambda} d\varphi \, e^{-\VRI{\varphi}{\varphi^o_{\Lambda^c}}}$.
\end{definition} 
In order to prove existence of Gibbs state related to this system of local specifications, the following probabilties $\Pi_\Lambda(F_\Delta|\varphi^o)$ for any bounded Borel sets $\Delta\subset \Lambda$, and any $F_\Delta\in \widetilde{ \mathcal {F}}_\Delta$, have to be controlled uniformly on $\Lambda\supset \Delta$ and $\varphi^o$.
By denoting $F_\Delta^{loc}$ the projection of $F_\Delta$ onto $\mathcal{F}_\Delta$, 
\begin{eqnarray}
\Pi_\Lambda(F_\Delta|\varphi^o)&=&\intConf{\Delta} d\varphi \mathbb{1}_{F_\Delta^{loc}}(\varphi)
\left(\frac{1}{\partFunc{\Lambda}{\varphi^o}}\intConf{\Lambda\setminus \Delta} d\psi e^{-\VRI{\varphi \cup\psi}{\varphi^o_{\Lambda^c}}}\right)\nonumber\\
&=& \intConf{\Delta} d\varphi \mathbb{1}_{F_\Delta^{loc}}(\varphi)
\frac{\corrFunc{\Delta,\Lambda}{\varphi}{\varphi^o}}{z^{\#\varphi}}\label{eq-SpecLoc}
\end{eqnarray} 
where $\corrFunc{\Delta,\Lambda}{\varphi}{\varphi^o}$, defined in~(\ref{eq-corrFunc}), play an important role in the expression of the Radon-Nikodym of the local specification $\Pi_\Lambda(\cdot|\varphi^o)$ with respect to some Poisson process in~$\Delta$.

Consequently, by combining the result of the proposition~\ref{prop-corrFunc} with $K=0$ and the equation~(\ref{eq-SpecLoc}) one derives for any bounded Borel sets $\Delta\subset \Lambda$, and any $F_\Delta\in \widetilde{ \mathcal {F}}_\Delta$ that,
\begin{eqnarray}
\Pi_\Lambda(F_\Delta|\varphi^o)&=&  \intConf{\Delta} d\varphi \mathbb{1}_{F_\Delta^{loc}}(\varphi)
\frac{\corrFunc{\Delta,\Lambda}{\varphi}{\varphi^o}}{z^{\#\varphi}}\nonumber\\
&\leq &  \intConf{\Delta} d\varphi \mathbb{1}_{F_\Delta^{loc}}(\varphi) \frac{\left(ze^{z|B(0,R)|}\right)^{\#\varphi}}{z^{\#\varphi}}\nonumber\\
&=& \intConf[ze^{z|B(0,R)|}]{\Delta} d\varphi \mathbb{1}_{F_\Delta^{loc}}(\varphi) \label{eq-Pi}
\end{eqnarray}
In particular, for the event $F_\Delta=[\Phi_{\Lambda}(\Delta)\geq m]$ one obtains~:
\[
P_\Lambda([\Phi_{\Lambda}(\Delta)\geq m]|\varphi^o)=\sum_{k=m}^{+\infty}\frac {z^k} {k!}\int_{\Delta^k} dx_1\ldots dx_k \corrFunc{\Delta,\Lambda}{\{x_1,\ldots,x_m\}}{\varphi^o}\leq \sum_{k=m}^{+\infty}\frac{(ze^{z|B(0,R)|}|\Delta|)^k}{k!}.
\]
In some sense, this means that $P_\Lambda$ is ``dominated" by the  non-normalized Poisson process with intensity $ze^{z|B(0,R)|} |\Delta|$.

In \cite{BBD3}, we proposed some simpler sufficient conditions based on the local energy in order to satisfy the Preston's theorem (\cite{Preston76} theorem~4.3 p.58) assumptions very useful for proving the existence of a stationary Gibbs state.

\begin{description}
\item[(LS) Local Stability:] there exists some constant $K\geq 0$~:
\begin{equation}\label{eq-LS}
 \VI{0}{\varphi} >-K,\quad \forall \varphi \in \Omega
\end{equation}
\item[(Q) Quasilocality:] for any bounded Borel set such that $0 \in \Delta$~: 
\begin{equation}
\left| \VI{ 0}{\varphi} -\VI{ 0}{\varphi _{\Delta}}\right| <\varepsilon (d(0,\Delta^{c})),\forall \varphi \in \Omega
\end{equation}
where $\varepsilon$ is a nonnegative decreasing function which vanishes asymptotically and $d(x,B)= \min\limits_{y\in B} d(x,y)$ is the Euclidean distance between a point $x$ and a Borel set~$B$.
\end{description}
The first assumption is not satisfied by our model based on the R-energy. Fortunately, one may replace it by the $R$-local stability assumption:
\begin{description}
\item [($R$-LS) $R$-local stability:] there exists some real value $R\geq 0$ such that $V$ is $R$-locally stable.
\end{description}
but also by the more general one based on the correlation function:
\begin{description}
\item [(UC) Upperbound of correlation function:] there exists some real value $\xi$ such that $\corrFunc{\Lambda}{\varphi}{\varphi^o}\leq \xi^{\#\varphi}$.
\end{description}

\begin{proposition} By assuming that $\Pi$ is a system of translation invariant local specifications based on some energy function satisfying \textbf{(UC)} (implied by \textbf{($R$-LS)}) and \textbf{(Q)},
the set $\mathcal{G}_0(\Pi)$ of stationary Gibbs measures is non empty.
\end{proposition} 
\proof The proof of this result is very similar to the one proposed in \cite{BBD3}. The only difference is that  condition \textbf{(3.7)} of theorem~4.3 (in \cite{Preston76}  p.58) is satisfied as a direct consequence of equation~(\ref{eq-Pi})
 \endproof

Consequently, we may assert the main result of this paper.
\begin{theorem}
The set $\mathcal{G}_0(\Pi^R)$ of stationary Gibbs measures associated to the system $\Pi^R$ of translation invariant local specifications based on the $R$-energy $\VRE$ defined in (\ref{def-VR}) is non empty.
\end{theorem}
\proof
$\VRE$ is $R$-locally stable (see proposition~\ref{prop-RLS}) and satisfies the finite range property (of proposition~\ref{prop-VR})
\endproof

We finally end this section by some concluding remarks.

\begin{remark} In the plane, the maximum number of Voronoi vertices (or equivalently, the number of Delaunay
triangles) is an upper bound for the number of holes (or more precisely, the Euler characteristic) generated 
by the Quermass-interaction model studied in~\cite{Kendall99}. Thus, there is a strong link 
between our model and the quermass-interaction model in the planar case when the grains are disks of fixed radius. The Quermass-interaction model~\cite{Kendall99} and nearest neighbours models~\cite{Baddeley89} defined using the Delaunay graph raised problems 
of stability in dimension greater than two
whereas  models presented here works in any dimension. 
\end{remark}

\begin{remark}
The assumption of relative compactness we discussed here for some particular nearest neighbours models is useful 
when we use correlation functions or local specifications. This assumption appears also in the modern large deviation theory 
and for having the sub levels of the specific entropy sequentially compact and existence of an accumulation 
point~\cite{Georgii93,Georgii94,Georgii96}.
\end{remark}

\begin{remark}
In~\cite{BBD5}, in order to study phase transition, we introduce the nearest-neighbour continuum Potts model where the soft repulsion between particles 
acts on a graph defined by the Delaunay edges. In order to prove the existence of this model, 
we simply add (in the spirit of~\cite{BBD4}) an hard-core component acting on all particles independently of their type. 
However, this result is still true without this assumption using the $R$-local Delaunay.       
\end{remark}

\begin{remark}
The stability of the finite energy and the temperedness of the mutual energy (see \cite{Ruelle69} p.32), implied by finite range property, provide results of \cite{Ruelle69} (p.41-58) concerning the existence of the pressure with free boundary condition and thermodynamic limit of microcanonical, canonical and grand canonical ensembles.
\end{remark}

\bibliographystyle{plain}
\bibliography{bbd}

\begin{thebibliography}{10}

\bibitem{Baddeley89}
A.J. Baddeley and J.~M{{\o}}ller.
\newblock Nearest-{N}eighbour {M}arkov {P}oint {P}rocesses and {R}andom {S}ets.
\newblock {\em Int. Statist. Rev.}, 57(2):89--121, 1989.

\bibitem{Baddeley96}
A.J. Baddeley, M.N.M. van Lieshout, and J.~M{\o}ller.
\newblock Markov properties of cluster processes.
\newblock {\em Adv. Appl. Prob.}, 28:346--355, 1996.

\bibitem{BBD4}
E.~Bertin, J.-M. Billiot, and R.~Drouilhet.
\newblock Existence of {D}elaunay {P}airwise {G}ibbs {P}oint {P}rocesses with
  {S}uperstable {C}omponent.
\newblock {\em {J}. of {S}tatist. {P}hysics}, 95:719--744, 1999.

\bibitem{BBD3}
E.~Bertin, J.-M. Billiot, and R.~Drouilhet.
\newblock Existence of ``{N}earest-{N}eighbour'' {G}ibbs {P}oint {M}odels.
\newblock {\em Adv. Appl. Prob.}, 31:895--909, 1999.

\bibitem{BBD2}
E.~Bertin, J.-M. Billiot, and R.~Drouilhet.
\newblock $k$-{N}earest-{N}eighbour {G}ibbs {P}oint {P}rocesses.
\newblock {\em Markov {P}rocesses and {R}elated {F}ields}, 5(2):219--234, 1999.

\bibitem{BBD5}
E.~Bertin, J.-M. Billiot, and R.~Drouilhet.
\newblock Phase {T}ransition in {N}earest-{N}eighbour {C}ontinuum {P}otts
  {M}odels.
\newblock {\em {J}. of {S}tatist. {P}hysics}, 114(1/2):79--100, 2004.

\bibitem{Connelly01}
R.~Connelly, K.~Rybnikov, and S.~Volkov.
\newblock Percolation of the loss of tension in an infinite triangular lattice.
\newblock {\em {J}. of {S}tatist. {P}hysics}, 105(1/2):143--171, 2001.

\bibitem{Georgii76}
H.-O. Georgii.
\newblock Canonical and {G}rand {C}anonical {G}ibbs {S}tates for {C}ontinuum
  {S}ystems.
\newblock {\em Commun. Math. Phys.}, 48:31--51, 1976.

\bibitem{Georgii94}
H.-O. Georgii.
\newblock Large deviations and the equivalence of ensembles for {G}ibbsian
  particle systems with superstable interaction.
\newblock {\em {P}rob. {T}heor. {R}elat. {F}ields}, 99:171--195, 1994.

\bibitem{Georgii96}
H.-O. Georgii and O.~H{\"a}ggstr{\"o}m.
\newblock Phase transition in continuum {P}otts models.
\newblock {\em Commun. Math. Phys.}, 181:507--528, 1996.

\bibitem{Georgii97}
H.-O. Georgii and T.~K{\"u}neth.
\newblock Stochastic {O}rder of {P}oint {P}rocesses.
\newblock {\em J. Appl. Prob.}, 34:868--881, 1997.

\bibitem{Georgii98}
H.-O. Georgii and V.A Zagrebnov.
\newblock On the interplay of magnetic and molecular forces in {C}urie-{W}eiss
  ferrofluid models.
\newblock {\em {J}. of {S}tatist. {P}hysics}, 93:79--107, 1998.

\bibitem{Georgii93}
H.-O. Georgii and H.~Zessin.
\newblock Large deviations and the the maximum entropy principle for marked
  point random fields.
\newblock {\em {P}rob. {T}heor. {R}elat. {F}ields}, 96:177--204, 1993.

\bibitem{Geyer94}
C.J. Geyer and J.~M{{\o}}ller.
\newblock Simulation {P}rocedures and {L}ikelihood {I}nference for {S}patial
  {P}oint {P}rocesses.
\newblock {\em Scand. {J}. of {S}tatist.}, 21:359--373, 1994.

\bibitem{Haggstrom00}
O.~H{\"a}ggstr{\"o}m.
\newblock Markov random fields and percolation on general graphs.
\newblock {\em Adv. Appl. Prob.}, 32:39--66, 2000.

\bibitem{Kendall90}
W.S. Kendall.
\newblock A spatial {M}arkov property for nearest-neighbour {M}arkov point
  processes.
\newblock {\em J. Applied Probability}, 28:767--778, 1990.

\bibitem{Kendall00}
W.S. Kendall and J.~M{\o}ller.
\newblock Perfect simulation using dominating processes on ordered spaces, with
  application to locally stable point processes.
\newblock {\em Adv. Appl. Prob.}, 32:844--865, 2000.

\bibitem{Kendall99}
W.S. Kendall, M.N.M. van Lieshout, and A.J. Baddeley.
\newblock Quermass-interaction processes: conditions for stability.
\newblock {\em Adv. Appl. Prob.}, 31:315--342, 1999.

\bibitem{Lebowitz72}
J.L. Lebowitz and E.H. Lieb.
\newblock Phase transition in continuum classical system with finite
  interactions.
\newblock {\em Phys. Lett. A}, 39:98--100, 1972.

\bibitem{Menshikov02}
M.~Menshikov, K.~Rybnikov, and S.~Volkov.
\newblock The {L}oss of {T}ension in an {I}nfinite {M}embrane with {H}oles
  {D}istributed according to a {P}oisson {L}aw.
\newblock {\em Adv. Appl. Prob.}, 34(2):292--312, 2002.

\bibitem{Moller98}
J.~M{{\o}}ller and R.P. Waagepetersen.
\newblock Markov connected component fields.
\newblock {\em Adv. Appl. Prob.}, 30:1--35, 1998.

\bibitem{Moraal76}
H.~Moraal.
\newblock The {K}irkwood-{S}alsburg equation and the {V}irial expension for
  many-body potentials.
\newblock {\em Physics Letters}, 59A(1):9--10, 1976.

\bibitem{Nguyen76}
X.X. Nguyen and H.~Zessin.
\newblock Integral and {D}ifferential {C}haracterizations of the {G}ibbs
  {P}rocess.
\newblock {\em Math. Nachr.}, 88:105--115, 1976.

\bibitem{Nguyen79}
X.X. Nguyen and H.~Zessin.
\newblock Ergodic theorems for {S}patial {P}rocess.
\newblock {\em Z. Wahrscheinlichkeitstheorie verw. Gebiete}, 48:133--158, 1979.

\bibitem{Preston76}
C.J. Preston.
\newblock {\em Random {F}ields}, volume 534.
\newblock Springer-Verlag, Berlin, Heidelberg, New York, 1976.

\bibitem{Ruelle69}
D.~Ruelle.
\newblock {\em Statistical {M}echanics}.
\newblock Benjamin, New York-Amsterdam, 1969.

\bibitem{Ruelle70}
D.~Ruelle.
\newblock Superstable interactions in classical statistical mechanics.
\newblock {\em Commun. Math. Phys.}, 18:127--159, 1970.

\end{thebibliography}

\end{document}